\documentclass[5p]{elsarticle} 
\usepackage{amsfonts}
\usepackage{textcomp}
\usepackage{amsmath}
\usepackage{bm}
\usepackage{xcolor}
\usepackage{hyperref}
\hypersetup{
    colorlinks,
    linkcolor={red!50!black},
    citecolor={blue!50!black},
    urlcolor={blue!80!black}
}
\usepackage{placeins} 

\usepackage{graphicx}
\usepackage{url}
\usepackage[utf8]{inputenc}

\journal{Nuclear Instruments and Methods in Physics Research Section A}

\usepackage[singlelinecheck=false]{caption}
\begin{document}
\begin{frontmatter}
\title {
Fast trajectory reconstruction techniques \\
for the large acceptance magnetic spectrometer VAMOS++
}

\author{A.~Lemasson\corref{cor1}} 
\ead{antoine.lemasson@ganil.fr}
\cortext[cor1]{Corresponding author}
\author{M.~Rejmund\corref{cor2}}

\address{GANIL, CEA/DRF - CNRS/IN2P3, Bd Henri Becquerel, BP 55027,
  F-14076 Caen Cedex 5, France}
\begin{abstract}

The large angular and momentum acceptance magnetic spectrometer VAMOS++, 
at GANIL, France, is frequently used for nuclear structure and reaction dynamics studies.
It provides an event-by-event  identification of heavy ions produced in nuclear reactions 
at beam energies around the Coulomb barrier. The highly non-linear ion optics of VAMOS++
requires the use of the heavy ion trajectory reconstruction methods in the spectrometer 
to obtain the high-resolution definition of the measured atomic mass number.
Three different trajectory reconstruction methods, developed and used for VAMOS++,
are presented in this work. The performances obtained, in terms of resolution
of reconstructed atomic mass number, are demonstrated and discussed 
using a single data-set of fission fragments detected in the spectrometer.


\end{abstract}
\end{frontmatter}

\section{Introduction}
\label{Sec:Int}

Magnetic spectrometers are very powerful tools to identify and characterize 
nuclei produced in nuclear collisions. In the recent years, large acceptance magnetic
spectrometers like VAMOS++~\cite{Pullanhiotan2008,Rejmund2011} at GANIL and 
PRISMA~\cite{Montagnoli2005} at LNL Legnaro 
have significantly contributed to the advances in the domain of nuclear structure and reaction 
mechanism studies.  They allow the isotopic identification and 
therefore the clean selection of reaction products on an event-by-event basis. 
The above-mentioned spectrometers are particularly suited for reaction at the beam energies 
near  the Coulomb barrier, where a large angular and momentum acceptance is necessary for efficient 
collection of the nuclear reaction products of interest.
However, a large angular and momentum acceptance is related to large sizes of optical elements,
composing the spectrometer, which often result in a design that 
exhibits a highly non-linear optics. Therefore, ray-tracing techniques for a 
trajectory reconstruction~\cite{Pullanhiotan2008,Rejmund2011}, are required, 
leading to the determination of the magnetic rigidity and related observables for 
the detected heavy ions. 

The isotopic identification, in terms of atomic mass number $A$ and atomic 
number $Z$, is the key issue as far as the magnetic spectrometers are concerned.
The basic measurement provided by a magnetic spectrometer is the magnetic rigidity 
$B\rho$ of the ion, it is proportional to the position in the dispersive plane
of the dipole magnet. The magnetic rigidity is related to the velocity $v$ and 
mass-over-charge ratio $(A/q)$, where $q$ is the atomic charge state.
The measurement of the velocity $v$ allows to obtain the corresponding 
mass-over-charge ratio $(A/q)$. The measurement of the atomic mass number 
$A$ requires the measurement of the atomic charge state $q$. Numerous spectrometers 
are used at high beam energies, from few tens to hundreds of MeV/A, so that the heavy 
ions of interest can be obtained in a fully stripped charge state, that is $q = Z$. 
Note that the energy necessary to reach the fully stripped charge state for a given atomic 
number $Z$ increases with $Z$. In this case the measurement of the correlation between the 
energy loss of the ion $\Delta E$ and its time-of-flight $t$ or velocity $v$, is often 
sufficient to determine the atomic charge state $q=Z$ and 
thus also the corresponding atomic mass number 
$A$~\cite{Bazin2003,Cortesi2020}. 
However, at energies near the Coulomb barrier, heavy ions have 
lower atomic charge state $q<Z$, with wide statistical distributions. 
In this case, the measurement of the total energy $E_{tot}$ of the incoming ion is 
necessary. The relation between the total energy $E_{tot}$ and atomic mass number $A$
combined with the measurement of magnetic rigidity $B\rho$ allows to determine the
atomic charge state $q$, for each ion individually. 
Due to the large angular and momentum acceptance, the measurement of the correlation
between the energy loss $\Delta E$ and the total energy $E_{tot}$ or velocity $v$
are necessary to obtain the atomic number $Z$~\cite{Kim2017}. Additionally, at lower 
beam energies, it is mandatory to minimize the effective thickness of every detector, 
through which the heavy ions are passing, the window thickness in particular, to reduce 
the related unmeasured energy losses.

VAMOS++ has been fruitfully used for a wide range of experiments covering (i)
particle and $\gamma$-ray spectroscopy using direct transfer~\cite{
        Obertelli2006, 
        Obertelli2006a,        
        Labiche2010, 
        Catford2010, 
        Fernandez-Dominguez2011,
        Brown2012,
        Flavigny2013,
        Assie2021,
        Galtarossa2021,
        Girard-Alcindor2022}, multi-nucleon transfer:~\cite{
        Rejmund2007, 
        Bhattacharyya2008, 
        Bhattacharyya2009, 
        Ljungvall2010,  
        Dijon2011, 
        Dijon2012,
        Celikovic2015,
        Klintefjord2017,
        Goldkuhle2019,
        Ralet2019,
        Ciemala2020,
        Goldkuhle2020,
        Siciliano2020,
        Ciemala2021,
        Siciliano2021,
        Ziliani2021,
        Perez-Vidal2022}, fission ~\cite{
        Shrivastava2009,   
        Alahari2014,    
        Navin2014, 
        Wang2015, 
        Biswas2016,
        Rejmund2016,
        Rejmund2016a,
        Kim2017,
        Hagen2017,
        Kim2017a,
        Navin2017,
        Dudouet2017,
        Hagen2018,
        Bhattacharyya2018,
        Delafosse2018,
        Singh2018,
        Dudouet2019,
        Biswas2019,
        Biswas2020,
        Banik2020,
        Wang2021,
        Rezynkina2022}, 
        and Coulomb excitation~\cite{Plaisir2014} (ii) fission~\cite{ 
    Caamano2013,
    Rodriguez2014,
    Farget2015,
    Caamano2015,
    Caamano2017,
    Ramos2018,
    Ramos2019,
    Ramos2019a,
    Ramos2020,
    Schmitt2021,
    Jhingan2022} and reaction dynamics~\cite{Benzoni2010, Golabek2010,  Watanabe2015, Marini2016,Fable2022,Fable2023}.

In this paper, three different trajectory reconstruction methods
elaborated for the VAMOS++ magnetic spectrometer throughout the years of operation
are described. The achieved performances will be discussed in terms of resolution 
of the reconstructed atomic mass number.

\section{VAMOS++ spectrometer}
\label{Sec:Spec}
\begin{figure}[ht]
\begin{center}
\includegraphics[width=1.0\columnwidth]{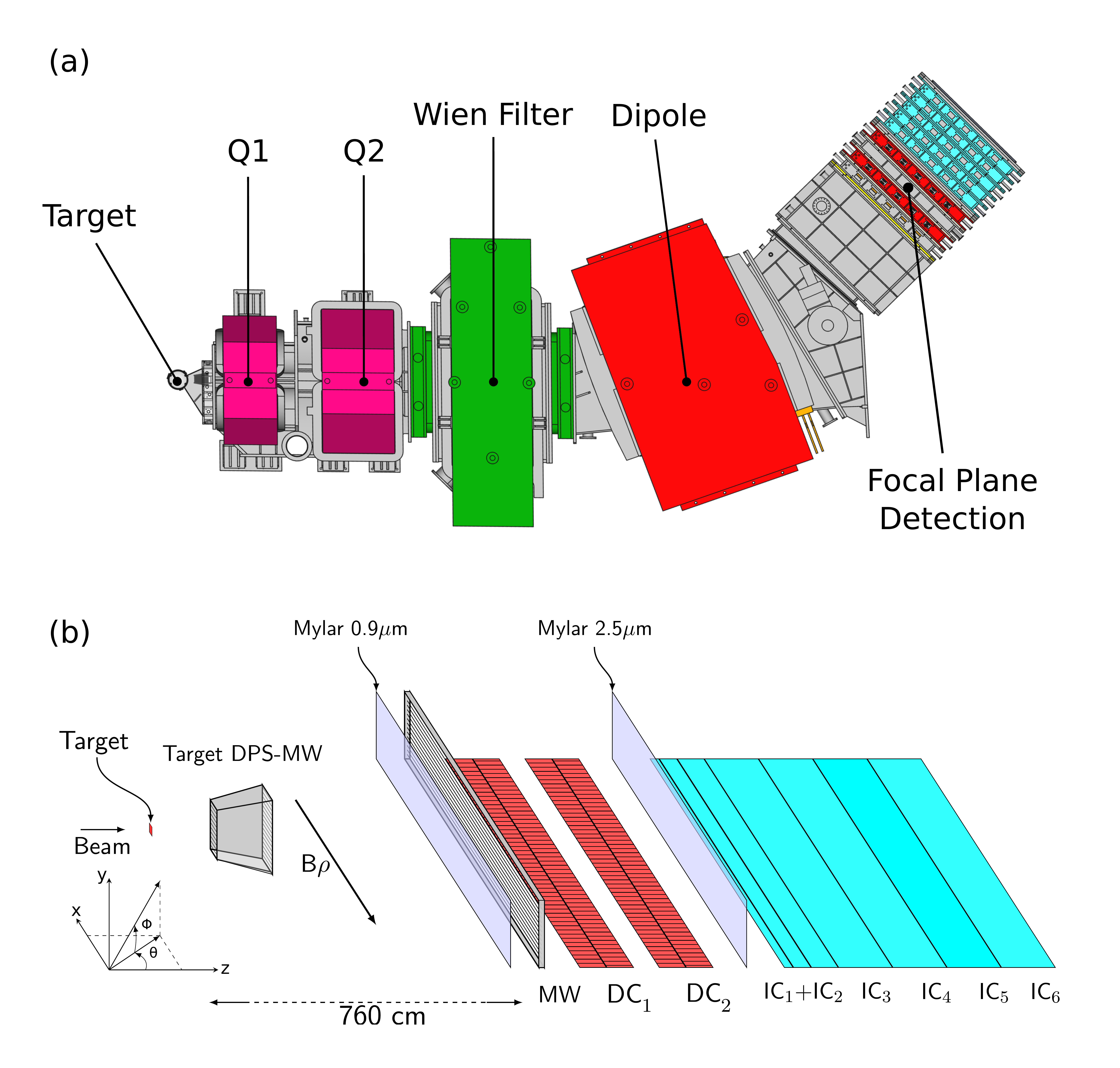}
\caption{\label{fig:1} 
(a)  Schematic view of the elements of the VAMOS++ spectrometer. The optical elements: two magnetic quadrupoles Q1 and Q2 for a vertical and horizontal focussing, Wien Filter WF (unused) and magnetic dipole D for horizontal dispersion.
(b) Schematic view of the VAMOS++ detection system. $\theta$ is the angle between the $z$-axis 
and the projection of the velocity vector of the trajectory on the $xz$ plane (symmetry plane), $\phi$ refers to the angle between the velocity vector and its projection on the $xz$ plane.
A dual position sensitive multi-wire proportional counter telescope (DPS-MWPC) is placed at the entrance of the spectrometer. At the focal plane, a multi wire proportional counter focal plane (MWPCFP), two drift chambers and segmented ionization chambers. Two Mylar windows used to isolate the detection gases are also shown.
}
\end{center}
\end{figure}

The layout of the VAMOS++ spectrometer is presented in Fig.~\ref{fig:1}(a). The optical elements of the VAMOS++ spectrometer~\cite{Pullanhiotan2008} consist of  two large aperture magnetic quadrupoles (Q1, Q2) for focusing heavy ions vertically and  horizontally, respectively and a magnetic dipole (D) for dispersing heavy ions horizontally.  The Wien Filter (WF) placed in-between the two magnetic quadrupoles and the magnetic dipole, is not used in the present work.

Due to the large angular and momentum acceptance of VAMOS++ the corresponding ion-optics 
is highly non-linear, as will be demonstrated in Sec.~\ref{Sec:Opt}, and a direct measurement 
of horizontal ion coordinate at the dispersive focal plane is insufficient to determine 
the magnetic rigidity $B\rho$ of the transmitted heavy ions. 
Therefore, more complex measurements of final coordinates, using the heavy ion tracking 
detectors, is necessary to perform ray-tracing and determine the magnetic rigidity $B\rho$, 
the velocity vector $\vec{v}$ and the trajectory length $l$. 
Typically, the focal plane detection system of VAMOS++ included two heavy ion tracking 
detectors, either secondary electron detectors (SeD)~\cite{Drouart2007,Pullanhiotan2008} or 
drift chambers~\cite{Pullanhiotan2008, Rejmund2011} , providing the measurement of horizontal 
and vertical coordinates  ($x_f$, $\theta_f$, $y_f$, $\phi_f$), where the index $f$ refers to final focal plane coordinates. Note, that the final focal plane coordinates were obtained on the image
focal plane placed $7600$~mm away from the target.
Additionally, an ionization chamber, either standalone~\cite{Kim2017} 
or combined with a plastic detector~\cite{Obertelli2006} or silicon wall
detectors~\cite{Pullanhiotan2008, Rejmund2011} were used used to provide 
the measurement of the energy loss and total energy correlation $\Delta E - E_{tot}$.
The stop for the time-of-flight measurement was provided by the plastic 
detector~\cite{Obertelli2006}, secondary electron detectors 
(SeD)~\cite{Drouart2007,Pullanhiotan2008} or multi-wire proportional counter
(MWPC)~\cite{Rejmund2011, Kim2017}. For the start for the time-of-flight ($t$) measurement,
during the early years of operation, the cyclotron radio-frequency was 
used~\cite{Pullanhiotan2008}. Since 2011, the multi-wire proportional counter,
at the entrance of VAMOS++,  (MWPC)~\cite{Rejmund2011} and since 2016 the dual 
position sensitive MWPC telescope (DPS-MWPC)~\cite{Vandebrouck2016} were used. 
The DPS-MWPC in addition 
to the timing signal provided the measurement of initial horizontal and vertical coordinates
at the target ($x_i$, $\theta_i$, $y_i$, $\phi_i$), where the index $i$ refers to initial 
coordinates at the target.

Schematic view of the VAMOS++ detection system, used to obtain the data presented 
in this paper~\cite{Kim2017}, is shown in Fig.~\ref{fig:1}(b).  The DPS-MWPC telescope was 
placed at the entrance of VAMOS++.
It consists of a pair of MWPCs, 
placed in a common gas 
volume. Each of MWPCs provides a precise timing signal as well as horizontal $(x)$ and vertical 
$(y)$ position. Therefore, a combination of all position measurements results in a 
determination of the interaction position at the target $(x_i, y_i)$ and the corresponding
scattering angles $(\theta_i, \phi_i)$.  The focal plane detection system of 
VAMOS++~\cite{Rejmund2011}, has an active area of $1000\times150$~mm$^2$ and consists of 
multi wire proportional counter focal plane (MWPCFP), 
two drift chambers and segmented ionization chamber. 
The MWPCFP, is 20-fold segmented in horizontal direction into sections 
and provides the timing signal for the measurement the time-of-flight ($t$). 
Each drift chamber measures the horizontal and vertical position that are used to obtain 
position $(x_f,y_f)$ and angles $(\theta_f, \phi_f)$ at the image focal plane.

The ionization chamber, is five-fold segmented in horizontal direction 
to improve its counting rate capabilities. It is also eight-fold segmented in depth. 
The atomic number $Z$ of the heavy ion is obtained from  the 
correlation of the measured $\Delta E$ and $E_{tot}$, see also Ref.~\cite{Kim2017}.  

The complete identification of the heavy ion using VAMOS++ spectrometer is obtained from the following relations:
\begin{align*}
\label{eq:eq1}
v & =   \frac{l}{t}, \beta = \frac{v}{c}, \gamma = \frac{1}{\sqrt{1 - \beta^2}} \\
(A/q)& =  \frac{B\rho}{3.107 \cdot \beta \cdot \gamma}\\
A_{(E_{tot}, \gamma)} & = \frac{E_{tot}}{931.494 \cdot (\gamma -1)} \\
q_{int} & =  \left\lfloor \dfrac{A_{(E_{tot}, \gamma)}}{(A/q)} + 0.5  \right\rfloor \\
A& = (A/q) \cdot q_{int}
\end{align*}
where
$v$ corresponds to the velocity in cm/ns, 
$l$ the trajectory length in cm,
$t$ the time-of-flight in ns,
$c$ the speed of light,
$B\rho$ the magnetic rigidity in Tm,
$E_{tot}$ the total energy in MeV,
$A_{(E_{tot}, \gamma)}$, is the atomic mass measured from total energy and velocity with resolution arising from the total energy resolution,  
$(A/q)$ the mass-over-charge ratio,
$q_{int}$ is the integer value of the atomic charge state and $\left\lfloor x + 0.5  \right\rfloor$ is the nearest integer value of x,
$A$ the atomic mass number.

\section{Ion-optics and ray tracing}
\label{Sec:Opt}

The trajectory of an ion along the spectrometer, can be described  using the standard ion optics
formalism using a six parameter vector $\vec{t} = (x,\theta,y,\phi,l,\delta)$ defined 
relative to a reference trajectory vector $\vec{t}_0=(0,0,0,0,l_0,1)$ for an ion with the 
reference magnetic rigidity $B\rho_0$.  The parameters $x$ and $y$ correspond to two 
transverse distances from the reference trajectory, see Fig.~\ref{fig:1}(b), $\theta$ is the 
angle between the $z$-axis and the projection of the velocity vector of the trajectory on the 
$xz$ plane (symmetry plane), $\phi$ refers to the angle between the velocity vector and its
projection on the $xz$ plane, $\delta = B\rho/{B\rho_0}$ defines the relative magnetic rigidity 
and $l$ is the path length from the target to the image focal plane.  
\begin{figure}[t]
\begin{center}
\includegraphics[width=1.0\columnwidth]{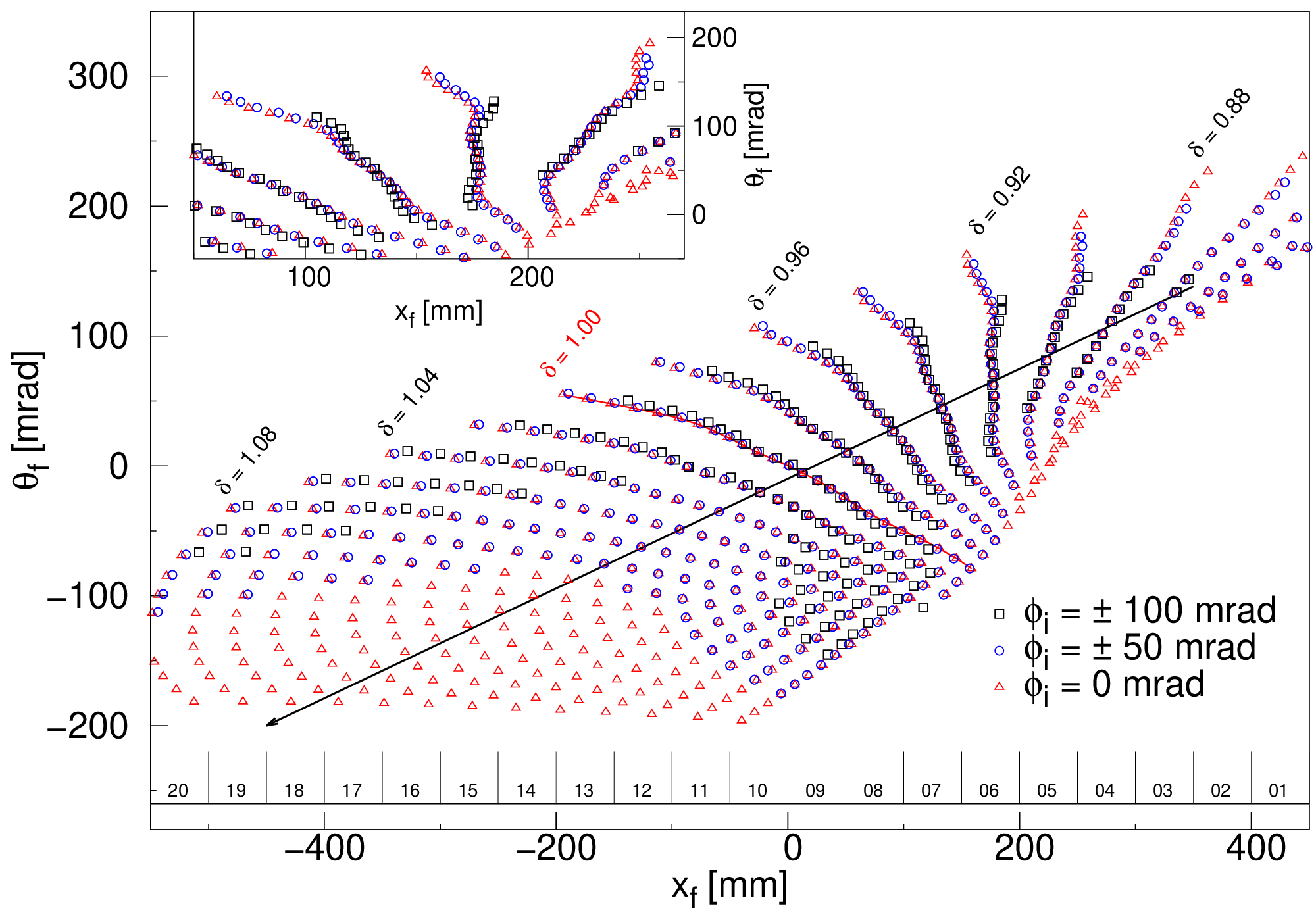}
\caption{\label{fig:2} Aberrations of the VAMOS++  spectrometer:
calculated angle ($\theta_f$) as a function of the positions ($x_f$) at the image focal plane of VAMOS++ illustrating  effects of aberrations in position and angle. 
The figure shows the final positions of the trajectories for discrete relative rigidity
($\delta = B\rho / B\rho_0$) varied in $2\%$ step and horizontal angle $\theta_i$ by $10$~mrad.  
The arrow shows the increasing B$\rho$ direction.
The trajectories, whose vertical angle $\phi_i = 0$~mrad are indicated by triangles, 
$\phi_i = \pm 50$~mrad by circles and $\phi_i = \pm 100$~mrad by squares.  
The arrow across the figure represents the positions of central
rays with $\theta_i = 0$~mrad. The inset shows a zoom in the region of low $B\rho$.
The horizontal position of the physical segments of the MWPCFP located in the focal 
plane is also indicated. 
}
\end{center}
\end{figure}

For ray-tracing purposes, a set of vectors in the initial phase space starting at the origin 
$x_i,y_i=0$ (assuming a point-like source) with well defined angles 
($\theta_{i_{k}},\phi_{i_{k}}$) and relative  magnetic rigidity 
($\delta_{i_{k}}$), $\vec{t}_{i_{k}} = (0,\theta_{i_{k}},0,\phi_{i_{k}},0,\delta_{i_{k}})$,  
were generated.  These vectors exceed a maximum angular and momentum
acceptance of the spectrometer and are defined within the range 
$\Delta \theta,\Delta \phi = \pm 160$~mrad
and $\Delta\delta=\pm 0.3$.  The generated density of initial vectors depends on the targeted 
resolution of the trajectory reconstruction ($\theta,\phi, \delta, l$) and the trajectory
reconstruction method applied (see below).
The trajectories of the ions in VAMOS++ were calculated using ray-tracing 
code ZGOUBI~\cite{Meot1999}. Realistic field descriptions for each magnetic element were 
considered in the calculation by incorporating median plane field maps issued from the 
3D field maps generated using electromagnetic computation code TOSCA~\cite{TOSCA}. 
The code ZGOUBI traces an ion through a system of magnetic fields and calculates the final
coordinates numerically by integrating the equation of motion in a magnetic field. 
The final coordinates of the trajectories are calculated on the image focal plane located 
$7600$~mm away from the target. For each trajectory, its initial trajectory vector trajectory,
$\vec{t}_i =(0,\theta_i,0,\phi_i,0,\delta_i)$ and 
final trajectory vector 
$\vec{t}_f = (x_f,\theta_f,y_f,\phi_f,l_f,\delta_f)$, 
are stored, provided that the trajectory reaches the image plane of VAMOS++. 
It should be noted that $\delta_i = \delta_f$.

The optics of the VAMOS spectrometer was designed to have the
horizontal plane as dispersive plane.  In the first order, the
dispersion is $\sim 2\text{cm}/\%$.  The horizontal coordinates in the
dispersive plane are dominantly decoupled and thus independent of the
vertical coordinates. This can be seen in the Fig.~\ref{fig:2}, where
the dependence on vertical coordinate ($\phi_{i}$) is limited. The
horizontal position of the beam at the target, in the first order,
results in a proportional position displacement in the focal plane
($\delta_{x_{i}} \propto \delta_{x_{f}} $). Therefore, in the first order, the horizontal size
of the beam spot of $\sim 2~mm$ would result in $\sim 1
\text{\textperthousand}$ uncertainty in the magnetic rigidity.

To illustrate the non-linearity in the ion-optics of VAMOS++ the
calculated $\theta_f$ as a function of $x_f$ in the image focal plane of VAMOS++ 
is shown in Fig.~\ref{fig:2}.  Overall effect of aberrations in position and angle can be seen
in the figure. The impact on the horizontal image, due to the non-null vertical coordinate 
$\phi_i$, is also indicated.  The triangles corresponding to the trajectories with 
$\phi_i=0$~mrad delimit a full $B\rho-\theta_i$ acceptance phase space of VAMOS++. 
The circles $\phi_i= \pm 50$~mrad and squares $\phi_i= \pm 100$~mrad indicate the  
complex 3-dimensional $(B\rho-\theta_i-\phi_i)$ acceptance phase space.
It is interesting to notice, that in the region of the low relative $B\rho$ 
\begin{itemize}
    \item in the first
order $B\rho$ is proportional to $x_f$, 
\item the ions with large $|\phi_i|$ are accepted 
for small $|\theta_i|$ and are progressively lost for larger $|\theta_i|$.
\end{itemize}
In the region of the high relative $B\rho$ 
\begin{itemize}
    \item  in the first order $B\rho$ is proportional 
to $\theta_f$,
    \item the ions with large $|\phi_i|$ are accepted for large $|\theta_i|$ 
and are progressively lost for smaller $|\theta_i|$.
\end{itemize}

Finally, the difference in position between the corresponding points $\phi_i=0,\pm 50, \pm 100$~mrad 
compared to the $B\rho$ step of $2\%$ illustrate the importance of the vertical coordinate in the 
determination of $\delta B\rho$ resolution in the few per mille limit.

\section{Trajectory reconstruction}
\label{Sec:Rec}

The goal of the trajectory reconstruction methods described in this
section is to provide the high resolution vector $\vec{t}_{rec} =
(\theta_i, \phi_i,\delta_i, l_f)$ from measured final or/and initial
coordinates on an even-by-event basis. Additionally, the
reconstruction methods should be implemented for efficient on-line
analysis of data.  In the following sections, three different
trajectory reconstruction methods will be described and their
performances, applied to the same experimental data set, will be
discussed.

\subsection{The Polynomial approach}
\label{Sec:Pol}

In the early years of operation of VAMOS++, the polynomial expansion method was used 
for the trajectory reconstruction.  
For each of the experimentally detected ions, the measured coordinates in the image focal plane
form a vector of final coordinates $\vec{t}_{exp} = (x_f,\theta_f,y_f,\phi_f)$.
Let $m=\,\theta_i,\,\phi_i,\,\delta_i, l_f$ denote four  coordinates to be reconstructed
forming a vector $\vec{t}_{rec} = (\theta_i, \phi_i,\delta_i, l_f)$.
Each of the coordinates $m$ can be expressed as independent non-linear function
$m = F_m(x_{f},\theta_{f},y_{f},\phi_{f})$.
It used only the final focal plane coordinates.  
Its $7^{th}$ order implementation is described in Ref.~\cite{Pullanhiotan2008}.  
This method was suitable for measurements involving the relatively light ions 
$A<70$~\cite{Rejmund2007, Bhattacharyya2008, Bhattacharyya2009}.
For heavier ions the $10^{th}$ order polynomial expansion method was introduced in $2008$. 
The $10^{th}$ order polynomial expansion method will be detailed in this section.

The four non-linear inverse transfer functions $F_m$ can be expressed as a $10^{th}$ order
polynomial of four variables $(x_f,\theta_f,y_f,\phi_f)$, measured exclusively in the image focal 
plane of VAMOS++. The inverse transfer function $F_m$ can be expressed as:
\begin{equation}
\label{eq:eq2}
\begin{array}{l}
   F_{m}=\displaystyle\sum_{i,j,k,l=0}^{i+j+k+l=10}C_{m_{ijkl}} 
    (x_{f})^{i} 
    (\theta_{f})^{j} 
    (y_{f})^{k} 
    (\phi_{f})^{l}
\end{array}
\end{equation}
where the coefficients $C_{m_{ijkl}}$ are related to the properties of the inverse transfer map of
the system.  The unknown coefficients $C_{m_{ijkl}}$ can be determined numerically. 
A computer program was developed for this purpose.
It uses the set of trajectories computed by ZGOUBI, as described in Sec.~\ref{Sec:Opt}, and
determines the best converged solution for $C_{m_{ijkl}}$ by fitting 
the polynomial expression in an iterative procedure.  
The initial trajectories have been calculated in steps of $d\theta_i = 10$~mrad, $d\phi_i = 10$~mrad and
$d\delta_i = 5\times10^{-3}$.
It should be noted that due to mid plane symmetry in the system, the coefficients $C_{{m}_{ijkl}}$ 
are null for $\theta_i$, $\delta_i$ and $l_f$  for odd values of $k+l$ and for 
$\phi_i$ for even values of $k+l$.
The remaining number of non-null coefficients for $\theta_i$, 
$\delta_i$ and $l_f$ is $511$ and for $\phi_i$ is $490$.  
Once the $C_{m_{ijkl}}$ coefficients were determined, they were used for the reconstruction
algorithm to map the measured final coordinates data on an event-by-event basis. Since the
algorithm is independent of any optics code once the coefficients are fixed, it can easily be
adopted in both on-line and offline event identification.  It should be noted that the VAMOS++
vertical magnification is of about $7$ in first order and that the finite vertical beam spot size 
induces a large uncertainty of the reconstructed values of $\theta_i$, $\delta_i$ and $l_f$. 
Therefore, the reconstruction of $\theta_i$, $\delta_i$ and $l_f$ were obtained solely from
$F_m(x_{f},\theta_{f},0,0)$ while the reconstruction of $\phi_i$ was obtained from $F_m(x_{f},\theta_{f},y_{f},\phi_{f})$.

\subsection{Two-dimensional (2D) mapping}
\label{Sec:2D}

The $10^{th}$ order polynomial approach, despite its complexity, is
unable to account for all the details of the aberrations with a single
set of coefficients for the whole focal plane.  This results in
reduced resolution of the reconstructed $\theta_i$, $\delta_i$ and
$l_f$. Attempts to overcome this limitation by sub-dividing the focal
plane phase space for the polynomial approach did not result in
significant improvement in the atomic mass resolution. Therefore, the
two-dimensional (2D) mapping approach was introduced in 2011 while
increasing the size of the focal plane detection~\cite{Rejmund2011}.

Let the coordinates $m = \theta_i, \delta_i, l_f$ be reconstructed using only the final image focal plane coordinates $x_f, \theta_f$ and the reconstruction of the coordinate $\phi_i$ remain as in the polynomial approach, Sec.~\ref{Sec:Pol}. 
The values of each of the $m$ coordinates can be stored in the two-dimensional array $M_m$,
as unsigned $2$-byte integer, with the numerical precision of $10^{-3}$,  $1$~mrad and $1$~mm for 
$\delta_i$ , $\theta_i$  and $l_f$, respectively. 
The dimensions of each array $M_m$ were chosen to be $1100 \times 550$ corresponding to 
steps of  $1$~mm and $1$~mrad for the coordinates  $x_f$ and  $\theta_f$, respectively.
Every of the arrays $M_m$ occupies $\sim 1.15$~Mb of RAM.
The  initial trajectories have been calculated in steps of 
$d\theta_i = 0.05$~mrad, $d\phi_i = 20$~mrad and 
$d\delta_i = 2\times10^{-4}$, for initial coordinates.
The step size have been chosen to ensure a continuity of the arrays $M_m$.
For each calculated trajectory the $\delta_i$, $\theta_i$ and $l_f$ coordinates  
were stored in the $M_m$ arrays, as a function of  $x_f$ and $\theta_f$ coordinates, 
as follows:
\begin{eqnarray*}
\label{eq:eq3}
k & = & \lfloor x_f + 600 + 0.5 \rfloor  \\
l & = & \lfloor \theta_f + 200 + 0.5 \rfloor \\
\\
M_{\delta_i}[k][l] &=& \lfloor \delta_i \times 1000 + 0.5 \rfloor \\
M_{\theta_i}[k][l] &=& \lfloor \theta_i + 200 + 0.5 \rfloor \\
M_{l_f}[k][l] &=& \lfloor l_f + 0.5 \rfloor
\end{eqnarray*}
where: $x_f$ and $l_f$ are in mm and $\theta_i$ and $\theta_f$ are in mrad.
The inverse procedure to obtain reconstructed parameters from measured ($x_f$, $\theta_f$) is straightforward.

\subsection{Four-dimensional (4D) mapping}
\label{Sec:4D}

The further improvement of the reconstruction quality was one of the
reasons to build the DPS-MWPC~\cite{Vandebrouck2016}, shown in
Fig.~\ref{fig:1}(b).  DPS-MWPC placed at the entrance of VAMOS++
provides the two-fold time measurement as well as two-fold vertical
and horizontal position leading thus the scattering angles $\theta_i,
\phi_i$ and the interaction point at the target $x_i, y_i$. Typical
resolutions of the reconstructed angles and positions on the target
were reported in Ref.~\cite{Vandebrouck2016} to be $\sigma=1.1~(1)$~mrad and
$\sigma=239~(30) \mu$m um respectively.

The four-dimensional (4D) mapping method, introduced in 2016,  is an extension of the 2D mapping,
where in addition the final image plane coordinates $x_f, \theta_f$, the initial
coordinates $\phi_i, \theta_i$ will also be used.
The reconstructed coordinates will be $m= \delta_i, l_f$.
The values of each of the m coordinates can be stored in the four-dimensional array
$M_m$, as unsigned $2$-byte integer, with a numerical precision of
$10^{-3}$ and $1$~mm for $\delta_i$ and $l_f$, respectively
The dimensions of each array $M_m$ were chosen to be $960 \times 450 \times 180 \times 260$,
steps of $1$~mm, $1$~mrad, $2$~mrad and $1$~mrad, for coordinates $x_f$, $\theta_f$, $\phi_i$ 
and $\theta_i$, respectively. Both arrays $M_m$ have a total memory requirement of $\sim 75$~Gb.
The initial trajectories have been calculated in steps of 
$d\theta_i = 0.05$~mrad, $d\phi_i = 2$~mrad and 
$d\delta_i = 1\times10^{-4}$, for initial coordinates.
The step size has been chosen such to guarantee a continuity of the arrays $M_m$. 
For each calculated trajectory the $\delta_i$ and $l_f$ were stored in the $M_m$ arrays
as a function of the $x_f$, $\theta_f$, $\phi_i$ and $\theta_i$ coordinates, as follows:

\begin{eqnarray*}
\label{eq:eq4}
k &=& \lfloor x_f + 550 + 0.5 \rfloor \\
l &=& \lfloor \theta_f + 200 + 0.5 \rfloor \\
m &=& \lfloor \phi_i/2 + 90 + 0.5 \rfloor \\
n &=& \lfloor \theta_i + 130 + 0.5 \rfloor \\
\\
M_{\delta_i}[k][l][m][n] &=& \lfloor \delta_i \times 1000 + 0.5 \rfloor \\
M_{l_f}[k][l][m][n] &=& \lfloor l_f + 0.5 \rfloor
\end{eqnarray*}
where: $x_f$ and $l_f$ are in mm and $\theta_i$, $\theta_f$ and $\phi_i$ are in mrad.
VAMOS++ acceptance phase space makes the $M_m$ arrays 
relatively sparse. The application of a zero suppression
algorithm, compressed row storage (CRS)~\cite{Barrett1994} allows to reduce 
the total required memory size from $\sim 75$~Gb
to $\sim 1$~Gb. To further facilitate the memory usage,
the compressed arrays can be stored in a binary format on disk and read 
into a permanent  shared memory segment made available for several analysis 
programs/processes at the same time.

It should be noted that, contrary to the polynomial and two-dimensional methods, 
the four-dimensional method does not ensure that every input vector ($x_f$,$\theta_f$,$\phi_i$,$\theta_i$) 
results in a valid $M_m$. This is due to the limited phase space of the spectrometer. 
This feature will be discussed in the following section and outlook.

\section{Experimental results}
In this section, the different methods are applied to a common benchmark 
experimental dataset and the performances in terms of reconstructed atomic mass are 
compared and discussed.
The experimental dataset arise from an experiment performed at GANIL, where the fission 
fragments were produced in fusion and transfer induced fission reactions using a 
$^{238}$U beam at the energy of $6.2$~MeV/u on a $^9$Be target ($1.6$ and $5$~$\mu$m thick).  
A schematic view of the experimental setup is shown in Fig.~\ref{fig:1}(a). 
The VAMOS++ spectrometer was placed at $20^\circ$ relative to the beam axis.
Further detail can be found in Ref.~\cite{Kim2017}.

\begin{figure}[ht]
\begin{center}
\includegraphics[width=0.97\columnwidth]{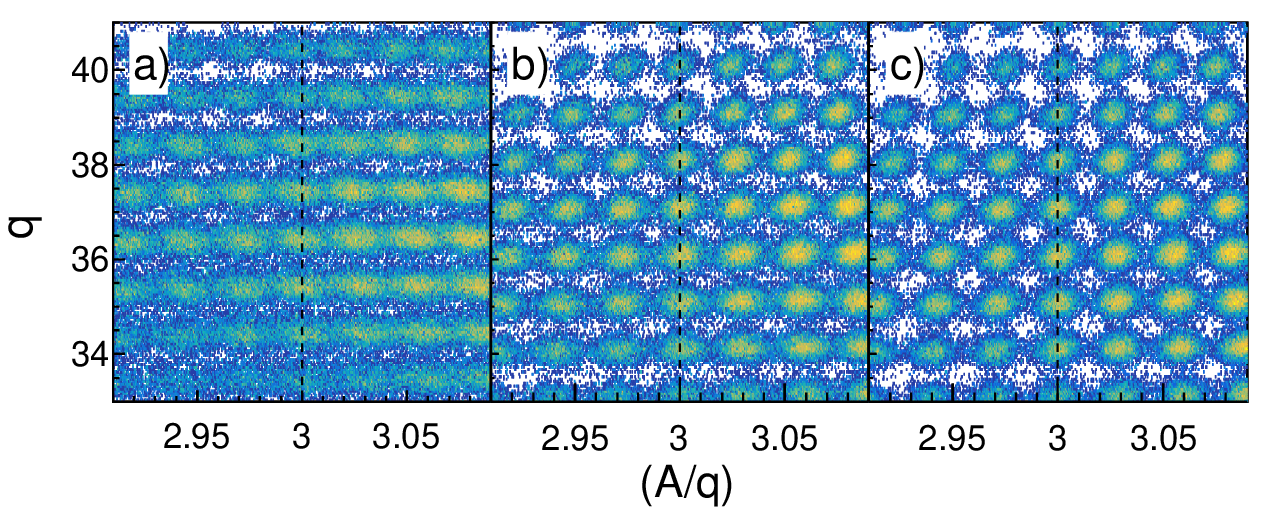}
\includegraphics[width=1.\columnwidth]{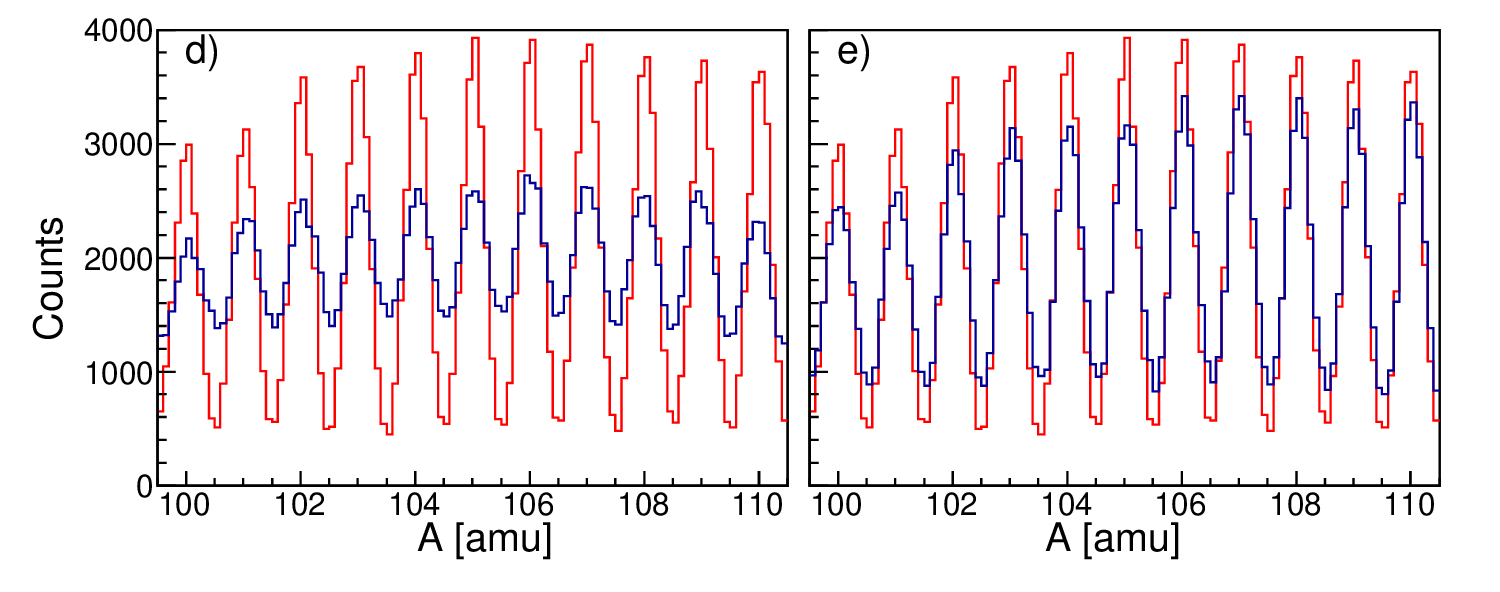}
\end{center}

\caption{\label{fig:3}
Atomic charge state $(q)$ as a function of mass-over charge ratio 
$(A/q)$ for physical section number $4$ of the MWPCFP ($ 250 mm < x_f < 300 mm $) using different reconstruction methods
a) polynomial approach, b) 2D mapping, c) 4D mapping.
The atomic mass number $A$ using 4D mapping in red compared to d) polynomial approach and 
e) 2D mapping methods in blue.
}

\end{figure}

The results will be first illustrated using the data  obtained from the physical section 
number $4$  of the MWPC detector of the focal plane. In Fig.~\ref{fig:3} the correlation between the atomic charge state $q$ and the mass-over-charge ratio $(A/q)$ is shown. Panels (a), (b) and (c) show the comparison of the results of the polynomial approach, the two-dimensional and four-dimensional mapping, respectively. The gradual improvement  of the quality of the identification can be seen. It should be noted that improved (A/q) reconstruction translates in an improved charge state resolution. 
The spectra of atomic mass number $A$ obtained from four-dimensional mapping method, in red, is compared to that obtained from polynomial approach in blue in Fig.~\ref{fig:3}(d) and to that obtained using two-dimensional mapping in blue in Fig.~\ref{fig:3}(e). It can be seen from the figures that atomic mass resolution ($\Delta A_{\scriptscriptstyle FWHM}/A$) is significantly improved by using the four-dimensional mapping. The resolution for $A=100$ is found to be  $5\text{\textperthousand}$ for the four-dimensional mapping compared to $7.5\text{\textperthousand}$ for the polynomial approach and $6\text{\textperthousand}$ for the two-dimensional mapping. This correspond respectively to 50\% and 20\% improvement of the atomic mass resolution for this region of the VAMOS++ focal plane. 
  
It can be seen in Fig.~\ref{fig:2} that the optical
  aberrations of VAMOS++ change as a function of the magnetic rigidity
  $B\rho$ of the ion and thus also as a function of the $x_f$ image
  plane coordinate. It has been observed that the differences between
  the results of different reconstruction methods are the largest in
  the region of low relative $B\rho$. In section number $10$ (see
  Fig.~\ref{fig:2}), polynomial approach and two-dimensional mapping
  were found to be equivalent and the four-dimensional resulted in a
  10\% improvement yielding to an atomic mass resolution of
  $6\text{\textperthousand}$ for $A=100$. In section number 16,
  the three approaches are found to give equivalent atomic mass
  resolution. This can be explained by the fact that for higher
  magnetic rigidity, the ions have in average higher velocity and the
  contribution of the time of flight resolution in the atomic mass
  resolution dominates.

\begin{figure}[!ht]
\begin{center}
\includegraphics[width=1\columnwidth]{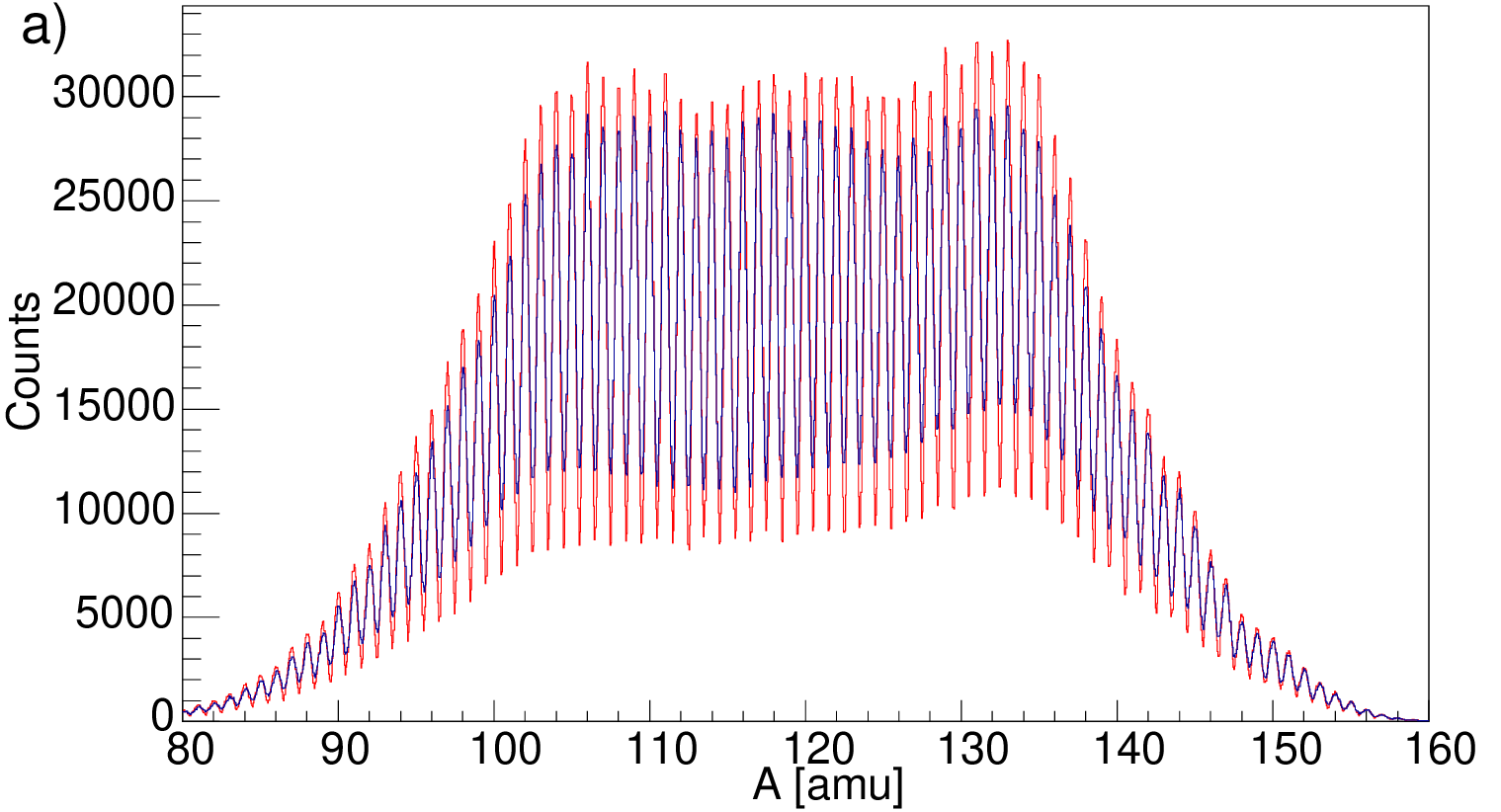}
\includegraphics[width=\columnwidth]{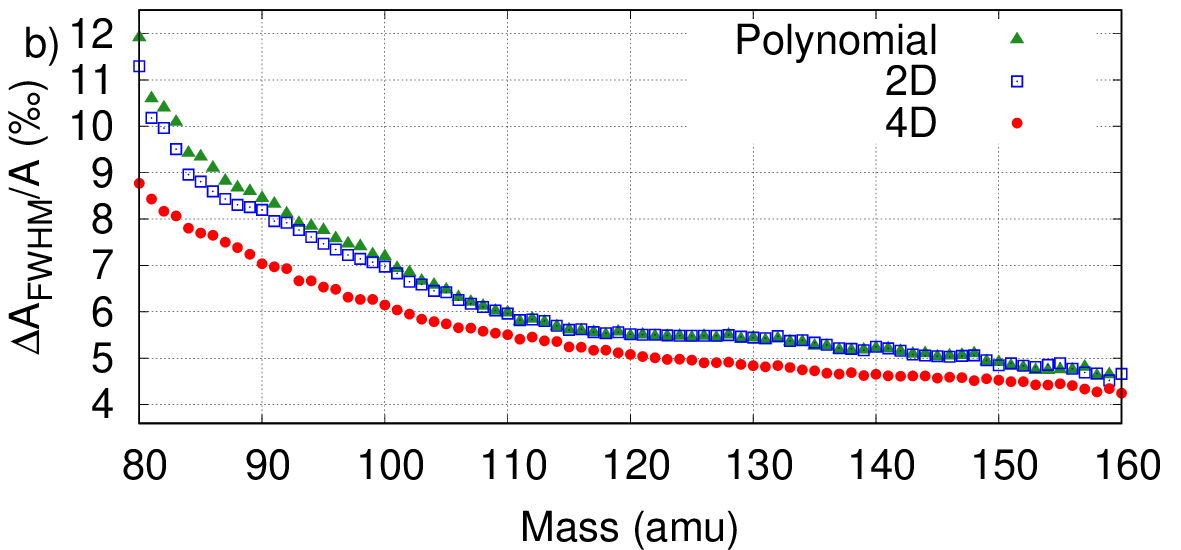}
\caption{\label{fig:4} a) Atomic mass number $A$ for events collected using a full image focal plane of VAMOS++, in red 4D mapping method and in blue 2D mapping method.
b) Resolution of atomic mass number $\Delta A_{\scriptscriptstyle FWHM}/A$ as a function of atomic mass number $A$ for the polynomial (green triangles), two- (open blue squares) and four-dimensional (red circles) reconstruction methods.}
\end{center}
\end{figure}

Further, the atomic mass number spectrum of fission fragments obtained for the complete focal plane is shown in Fig.~\ref{fig:4}(a). The spectrum shown in red correspond to the four-dimensional  mapping reconstruction and the spectrum shown in blue to that of the two-dimensional mapping. Figure~\ref{fig:4}(b) shows the associated FWHM resolution obtained for atomic mass number $\Delta A_{\scriptscriptstyle FWHM}/A$, in per mille (\textperthousand), as a function of $A$. A clear improvement in width $\Delta A_{\scriptscriptstyle FWHM} $ of about $8\%$ for the four-dimensional mapping method, relative to other methods, can be seen in the figure.  The downward slope of the $\Delta A_{\scriptscriptstyle FWHM}/A$ as a function of increasing $A$ results from nearly constant $\Delta A_{\scriptscriptstyle FWHM} $.

Finally, the efficiency of the reconstruction method is 100\% for polynomial and two-dimensional methods. However, in the case of the four-dimensional method, a typical efficiency of $\sim 97$\% was obtained. This reduced efficiency can be related to the highly constrained phase space arising from the optics of the spectrometer. As an example, an extension of the beam on target (typically $\Delta x_{\scriptscriptstyle FWHM} = 1.2$~mm and $\Delta y_{\scriptscriptstyle \scriptscriptstyle FWHM} = 1.5$~mm~\cite{Vandebrouck2016}) will result in some initial angles ($\theta_i$, $\phi_i$) residing out of the calculated phase space assuming a point-like beam spot ($x_i = 0$ and $y_i = 0$). As a consequence, for such cases, the four-dimensional method will not provide reconstructed $\delta_i$ and $l_f$. Nevertheless, the two-dimensional method can be used as a failover solution for these events at the cost of a reduced resolution. The implementation of the reconstruction method accounting for a size of the beam spot is considered for future work. It should be however noted that, the use of such a method on an event-by-event basis, will require an increase of the matrix size at least by a factor of $\sim 10$. 

\section{Summary and perspectives}
The large angular and momentum acceptance magnetic spectrometer VAMOS++, 
is particularly well suited for the studies of nuclear structure and reactions dynamics
at the beam energies near the Coulomb barrier.
The main objective of VAMOS++ is to provide on an event-by-event basis the isotopic
identification of the reaction products of interest. A high-resolution of the atomic
mass measurement requires the use of the trajectory reconstruction methods, due to 
the highly non-linear ion optics of the spectrometer. Three trajectory reconstruction 
methods have been developed and used in the past years 
\begin{enumerate}
    \item polynomial approach, using  $x_f$, $\theta_f$, $y_f$ and $\phi_f$ coordinates,
    \item two-dimensional (2D) mapping using $x_f$, $\theta_f$ coordinates,
    \item four-dimensional (4D) mapping $x_f$, $\theta_f$ and  $\phi_i$ and  $\theta_i$ coordinates.
\end{enumerate}

These methods makes use of the parameters derived from of set of trajectories of the ions in VAMOS++ 
calculated using ray-tracing code ZGOUBI~\cite{Meot1999}.
All methods are fast and allow an efficient treatment of the experimental data in an on-line and 
off-line analysis. The trajectory reconstruction method were applied to a single data-set
of fission fragments. 

An improvement of about $8$\% was obtained using 4D mapping method as compared to the polynomial approach and two-dimensional mapping method, leading to $\Delta A_{\scriptscriptstyle FWHM}/A$ ranging from $4.5$\textperthousand\space  
($\frac{1}{220}$) for heaviest fragments to  $9$\textperthousand\space ($\frac{1}{110}$) for lightest fragments.

The trajectory reconstruction methods presented on this work are based on the assumption that
the beam spot size at the target is point-like, $x_i = 0$ and $y_i = 0$,
while the typical beam spot size is $\Delta x_{\scriptscriptstyle FWHM} = 1.2$~mm and
$\Delta y_{\scriptscriptstyle \scriptscriptstyle FWHM} = 1.5$~mm~\cite{Vandebrouck2016}. 
In future work, it is foreseen to extend the mapping method including 
the event-by-event  measurement of the interaction point on the 
target and investigate its impact of the resolution of the reconstructed 
atomic mass number.

\section*{Acknowledgements}
The authors thank B.~Jacquot and D.~Ramos for numerous fruitful discussions.

\section*{References}
\bibliographystyle{elsarticle-num2} 
\biboptions{sort&compress}
\bibliography{NIM_VRec_Rev}

\end{document}